\title{Decision Taking versus Promise Issuing}
\author{Jan A. Bergstra\\Informatics Institute, University of Amsterdam\footnote{%
Science Park 904, 1098 XH Amsterdam, The Netherlands}
}

\documentclass[12pt]{article}

\date{}

\usepackage{alltt}
\usepackage{epsfig}
\usepackage{a4}
\usepackage{times}

\def\beq{\begin{eqnarray}}
\def\eeq{\end{eqnarray}}
\def\2{\frac{1}{2}}

\newtheorem{definition}{Term}

\begin{document}
\newcommand{\promise}[1]{\stackrel{#1}{\longrightarrow}}

\maketitle

\begin{abstract}
An alignment is developed between the terminology of outcome oriented decision taking and
a terminology for promise issuing. Differences and correspondences are investigated between the
concepts of decision and promise.

For decision taking, two forms are distinguished: the external outcome delivering form and internalized decision taking.
Internalized decision taking is brought in connection with Marc Slors' theory of self-programming. 

Examples are produced for decisions and promises in four different several settings each connected with software technology: instruction sequence effectuation, 
informational money transfer, budget announcement, and division by zero.
\end{abstract}

\section{Introduction}
Decisions and promises are interconnected notions, both of importance in a range of areas inside and outside technology.
Mark Burgess has initiated a sustained investigation to the application of promises in information technology and the
author has taken on board the challenge to develop at theory of promise taking that can be 
combined in a modular fashion with other theories about human and artificial action and cognition. 

For Burgess the key
observation was that promises constitute the preferred form of communication by autonomous agents, thereby
creating a significant distance from obligation based accounts of promising. For myself
the key observation was that a decision is an action rather than the result of an action and that the outcome of a decision
must be distinguished from the consequences of putting that outcome into effect. That has led me to the concept 
what I have termed outcome oriented decision taking.

If agents taking notice of a decision outcome are assumed to be autonomous in acting accordingly, 
autonomy is as much an issue for decision taking as it is for promising.\footnote{%
The role of autonomy for agents acting upon the reception of decision outcomes has not been brought into focus in my 
work \cite{Bergstra2011a,Bergstra2012b,Bergstra2012c,Bergstra2012d} on outcome oriented
decision taking and in these papers the possibility that decision outcomes imply obligations for other agents is left open.}

In this paper I will bring together decisions and promises in a single framework. The framework is extended with
so-called internalized outcome oriented decision taking drawing from work on the philosophy of cognition by Mark Slors.
An extensive collection of examples of promises and decisions, including promises about decisions and decisions about promises,
is presented with the aim to demonstrate the expressive power and flexibility of these concepts when applied to
seemingly trivial cases.

\section{Terminology for promise issuing}
This section presents an alignment of the terminology of promising  that has recently been developed by Mark Burgess in
\cite{Burgess2005,Burgess2007,BurgessF2007}, consistent with \cite{Stoljar1952}, and subsequently worked out in 
some more detail in \cite{BergstraB2008}
 to the terminology of decision taking in style of
OODT (outcome oriented decision taking) that has been developed in 
\cite{Bergstra2011a,Bergstra2012b,Bergstra2012c,Bergstra2012d}. I will assume that the 
reader is familiar with the account of promises that is developed in \cite{Bergstra2011a}.

\subsection{Deviations from ``static theory of promises'' terminology.}
Five deviations from the SToP (static theory of promises) terminology that was 
set out in \cite{BergstraB2008} on the basis of \cite{Burgess2005,Burgess2007,BurgessF2007} will be made: 
\begin{enumerate}
\item I will speak of promise issuing where SToP uses promise making, 
\item Promise making in turn will be used to refer to a larger process covering several phases including a promise preparation
phase, optionally containing a promise issuing thread design phase, and ending with a promise issuing phase that
consists of the effectuation of the promise issuing thread. In SToP there is no term for the whole that I propose to
 denote with promise making.
\item Promise is understood as the act of promise issuing, rather than as its outcome.
\item The promise body of SToP reappears as a promise outcome understood as a conceptual unifier of
the plurality of promise outcome instances.\footnote{%
A promise outcome will consist of as many copies of (representations of) the promise body (in SToP terminology) 
as there are 
agents in the scope of the promise.}
\item In SToP a promise (now issuing a promise) consists of documenting an intention. 
Documenting suggest the production of a physical document, but this connotation is in fact too strong. I prefer to state that
promise issuing consists of expressing an intention, whereas a promise outcome consists of the complementary impressions that this 
expression has created with, or within, the agents in scope of the promise. This formulation covers the case
of a promisee listening to or looking at a promiser, without any physical document external to the agents involved playing a role.
\end{enumerate}
These modifications of promise terminology can easily be accommodated in SToP by modifying its terminology. 
A survey of that adaptation is given in the Appendix.
The sole reason for proposing these changes in terminology stem 
from the objective to find an alignment between promise theory (with modified terminology) 
and the framework of outcome oriented decision taking.

\subsection{Promise making}In OODT decision making is a process involving decision 
preparation and culminating in decision taking that 
produces a decision outcome. In strict alignment with OODT I propose that promise making is a process 
that involves promise (outcome) preparation, and which 
may include promise issuing, thereupon resulting in a promise outcome. 

\subsubsection{Promise making thread and startup of that thread}
Promise making can be a complex matter and depending on the kind of promise to be made a dedicated thread for
making that kind of promise may be designed. Starting up such a thread may itself require taking a 
decision to that end.\footnote{This description follows the pattern that was described for decision making threads and
decision taking for decision making thread startup in \cite{Bergstra2012d}. A promise making thread may be effectuated
concurrently by a single agent by way of strategically interleaved multi-threading (see \cite{BergstraM2007}).}

Promise making can have a shared part from which a plurality of promises can be generated, but it may also have 
arts specific for a single promise issuing.
\subsubsection{Promise preparation}
Complex promises must be thought out before being issued. Promise preparation refers to that part of the
promise making process in which a promise outcome (or rather the content of a forthcoming promise outcome)
is designed.\footnote{%
Promise preparation is the process that leads to the preparation of a promise body in the 
terminology of \cite{BergstraB2008}.
Promise preparation takes place in advance of promise issuing event.}

In complex cases promise preparation is explicitly structured
and takes place during the effectuation of a dedicated promise making thread.

\subsection{Promise issuing}
I will follow the definition of a promise as given in SToP (\cite{BergstraB2008}). Deviating from that paper, however, 
I will say that a 
promise is issued instead of being made. Admittedly by speaking of promise issuing I also deviate from conventional language. 
This deviation has been chosen in order to have a better alignment with the theory of decision taking
and decision making that was formulated in \cite{Bergstra2011a,Bergstra2012b,Bergstra2012c}. 
The approach to decisions of those papers is referred to as 
outcome oriented decision taking (OODT). In OODT decision taking is an action of deciding. Similarly I maintain that
promise issuing is an act of promising. Whereas decision taking produces a decision outcome, 
promise issuing produces a promise outcome.
 
In particular, issuing a promise consists of (i) the production of a promise outcome by the promiser,\footnote{%
The production of a promise outcome may employ a design that emerged from the preparation phase.
In computer programming terminology the promise may be build from its design.}
(ii) the delivery of that promise outcome to the promisee, and (iii) revealing the promise outcome to
the agents in scope of the promise.

I will not distinguish between a promise outcome and its incarnations in the minds of the promisee or of 
other agents in scope of the promise. I will maintain that a promise outcome consists of a 
plurality of promise outcome instances
that each have been taken on board by an agent in scope of the promise. Considering the instances real and the
promise outcome merely a conceptual common core of the plurality of ``actual'' instances, the notion that issuing 
a promise comprises producing a promise outcome must not be misunderstood in the way that the promise outcome
must by necessity have an independent physical existence or representation from which its instances are subsequently 
generated or derived.

\subsubsection{Implied promises}
When a promise is issued other promises may be either explicitly or implicitly implied. For instance if A promises B to drive
B with A's car to the airport ``now'', then implicitly A also promises to be in the possession of a driving license and
to have the license at hand. Such promises are implied promises given the original promise. 

If A promises B that A will pay for a meal in a restaurant then in most cases A will subsequently and silently (see \ref{SilentPr}) 
issue the promise to restaurant staff that A  will pay for what has been ordered.\footnote{%
In \cite{Bergstra2012d} implied decisions were introduced. Implied decisions, understood as decisions that must be taken after,
and because of, a certain decision was taken, are almost unavoidable if the first decision is 
about an activity above some threshold complexity. The notion of an implied promise is slightly different in that implied promises
are likely to be issued silently and, when made explicit, to occur concurrently  with the initial promise.}

\subsubsection{Promise issuing thread}
Promise issuing comprises the creation of promise instances in the minds of a 
promisee and possibly of other agents in scope of the process. This can be achieved in many different ways, and from
a certain complexity onwards the action itself becomes a composition of successive and perhaps also concurrent 
atomic parts that is best seen as the effectuation of a (multi-)thread that has been 
explicitly designed for the purpose of issuing a particular kind of promise in a particular 
context.

Designing a dedicated promise issuing thread may be considered part of the promise preparation phase of 
a promise making process, at least  if the case is new or unique. 
In other cases during promise preparation a known and probed design for a promise issuing thread may be 
retrieved from a library containing such information.

\subsection{Promise outcomes as distributed information items}
\label{reasoning5}
Promise outcomes are entities that either are installed independently in the minds
of one ore more agents in scope of the promise or otherwise  are posted on a blackboard accessible to the 
agents in scope. 

Promise outcomes are distributed but coherent collections of information items, 
each individual item being considered an instance of the promise outcome.
In functional terms promise outcomes and decision outcomes each serve as inputs for subsequent
cognitive processes. In other words: promise outcomes and decision outcomes are inputs for reasoning processes.

If A issues a promise (with body) P to promisee B with (besides B) agents \{B$_1$,...,B$_n$\} in scope, then
after that event all agents in \{B, B$_1$,...,B$_n$\} possess some occurrence of (an instance) of P. Thus promise issuing 
may lead to a distributed outcome. Each of these receiving agents will use its own copy of the promise body for
reasoning in different ways. 

For an agent B$^{\prime} \in$ \{B, B$_1$,...,B$_n$\} 
receiving (or rather incorporating) its own instance (or occurrence, or even incarnation) of the promise outcome, written
$\rho$(P,B$^{\prime}$),
marks the start of the following five concurrent and mutually interacting reasoning processes:
\begin{enumerate}
\item Reasoning towards the determination of expectations (held by B$^{\prime}$) that can be derived from  
 $\rho$(P,B$^{\prime}$). 
(Promise outcome based expectation creation.)
\item Reasoning about which observations support an assessment of the degree to which A is keeping or has kept its promise. 
(Promise compliance assessment method generation.) 
\item Given the available assessment methods (which may be updated on the fly): 
promise (keeping) assessment is the process that determines the degree
of promise keeping regarding $\rho$(P,B$^{\prime}$) that is shown by A from the perspective of B$^{\prime}$. Promise assessment
produces feedback in terms of a possible update of B$^{\prime}$'s trust in A. (Promise assessment by 
promise assessment method application.)
\item Reasoning about the amount of trust that B$^{\prime}$ has in A given what B$^{\prime}$ 
concludes about A's keeping promise $\rho$(P,B$^{\prime}$). (Promiser trust maintenance.)
\item Reasoning about what to do next, given that B$^{\prime}$ has received 
 $\rho$(P,B$^{\prime}$). (Promise outcome based planning and plan adaptation.)
\end{enumerate}

The simultaneous generation of these different but interrelated reasoning processes for a plurality of agents constitutes
a remarkable expressive power that justifies the concept of promising in a  setting of informaticology 
(see \cite{Bergstra2012f}).

\subsubsection{Promise outcome instance erosion}
The various instances of a promise outcome have a limited shelf life. 
When based in an agent's mind the instance may become forgotten or twisted. 
When posted somewhere it may loose its ``currency'' and may eventually be dropped. 
Different promise outcome instances (for the same promise) are likely to erode with different speeds.

\section{Internalized outcome oriented decision taking (IOODT)}
OODT assumes that observable action and fact marks the appearance of a decision outcome. Thus if a decision taker A
produces a decision that decision in its turn produces an outcome which is external to the decision taker and which is
in most cases observable for other agents. 
After noticing and observing the decision outcome and in view of the known role of A these other agents may behave
in such a way that the decision outcome is put into effect.\footnote{%
Effectuation of a decision outcome is not considered part of the decision making process.}
 
I will distinguish external OODT (EOODT) and internalized OODT (IOODT). Internalized OODT does without
a decision outcome external to the decision taker. Instead the (internalized) decision outcome 
resides inside the mind of the 
deciding agent for some period of time. OODT as proposed in \cite{Bergstra2011a,Bergstra2012b,Bergstra2012c,Bergstra2012d}
 is considered to be both external and explicit, and henceforth
I will assume that OODT refers to EOODT by default.
Thus EOODT may be considered to represent a subclass of decision taking phenomena only, 
while another (possibly hypothetical) class  of decision taking occurs entirely inside the (body or) 
mind of the decision taker. 

 IOODT must be distinguished from 
SAOODT (single agent OODT) which is a common instance of EOODT. 
In SAOODT a single agent serves as a decision taker 
alone and by deciding that agent singlehandedly  produces the corresponding (external) decision outcome.

It is reasonable to assume that in a significant proportion of cases of SAOODT (say with A as a deciding agent),
 the decision making process contains a phase corresponding to an instance of IOODT enacted by 
A which then serves as a cause for subsequent EOODT. The causation of subsequent EOODT from an
internalized decision outcome will  be unconsciously  triggered from within B. Below I will consider the case that the 
deciding agent is human only.\footnote{%
For robotic agents there engineers avail of a freedom of design that makes these issues less interesting for
principled investigation. At the same time doing experimental work on robotic consciousness still seems  
to be way ahead.}

\subsection{Adapting Marc Slors' theory of programs for the mind}
IOODT fits perfectly well with Marc Slors' theory of willing and human agency as it has been 
put forward in \cite{Slors2012,Slors2013}. In that
work Slors surveys known worries coming from the psychology of cognition and in particular from neuroimaging based
brain science  that conscious willing cannot conceivably be the cause of an action. These worries arise
because in a range of by now classical observational studies it has been observed that
phenomena in the brain of an agent which demonstrably predict the action temporally precede the 
conscious awareness of willing that action. 

It must be accepted that most if not all action performed by an agent is caused by phenomena or by states of affairs 
that are unconscious to the agent involved. Slors claims that this can't be the whole story 
because if fails to account for deliberately planned activity.

Slors suggests that by willing to carry out a plan an agent constructs a program to that extent. That program
is loaded in the brain of the agent and it resides ``in brain'' in that form waiting until an 
unconscious event triggers its effectuation. The causal chain from willing
to doing is of an indirect nature and this viewpoint rescues both the relevance of willing, 
that has come under philosophical fire from the side of neuroscientific experts, and the proven absence of a conscious
cause for an action immediately preceding that action, 
an absence which has been established by neuroscientific research beyond reasonable doubt so it seems.

I follow Slors' attractive account of willing and what it produces in the related case of decision taking with one modification: 
I prefer to speak of control code rather than of a program. (Computer) control code\footnote{%
For a definition of control code see \cite{BergstraM2009}.}
 is a notion drawn from computing just as the concept of a (computer) program. 
 However, ``control code'' is more abstract than ``program'' in the sense that when
speaking of control code one does not presuppose a mechanical understanding of how it is put into effect. 
For a computer program that understanding is commonly (or even by definition) supposed to exist.\footnote{%
That perspective on programs has been worked out in detail in the program algebra of \cite{BergstraL2002}.}

\subsubsection{Requirements on IOODT decision outcomes}
The decision outcome of an IOODT decision taking event performed by agent A 
will be a control code (hereafter named IDOCC for internalized decision outcome control code) 
that is located in the brain of A. That
control code can be put into effect by an unconscious (for A) trigger that has been 
activated within the decision taker's brain. 
It may be considered a role of A that it can place an IDOCC firmly in
A's brain.\footnote{%
In OODT (see \cite{Bergstra2011a}) the role of a deciding agent constitutes an essential aspect of decision taking.}

For an IDOCC to serve as a decision outcome of an act of decision taking 
it matters that  an IDOCC cannot be easily modified by the agent. 
Once loaded an IDOCC rests in place until its activation is
triggered by some event. An IDOCC can be deactivated when the agent takes another decision to that extent.

Secondly it is important for the status of an IDOCC as a decision outcome that its effectuation, once triggered, is not
dependent on A's memory of earlier having taken the decision that created the IDOCC. The effectuation of an IDOCC may 
also not depend on other residues in the brain of A of processes that led to the decision from which an IDOCC is
considered an outcome.

\subsubsection{IOODT  versus action determination and versus real time choice}
Action determination (AD) as specified in \cite{Bergstra2012c} refers to the determination of a course of action by A in real time.
AD differs from IOODT by the absence of an intermediate stage when the control code for choosing in real time 
has been produced and loaded.
AD need not involve a conscious phase in which the agent is willing to act in certain ways, and in which
by means of willing or of decision taking a decision outcome is inscripted in a writeable part of the 
mind of the deciding agent. 

Real time choice (RTC) occurs if an agent performs a choice between different options or utterances
and when the mechanism for choosing has not been predetermined by a previous decision (IOODT style). RTC may
be considered a special case of AD.

\subsubsection{IOODT: production and deployment of control code for the mind}
Taking the analogue between computer control code generation and internalized decision taking a bit further
one may say that internalized (outcome oriented) decision taking is implemented inside the brain of a
human agent in ways comparable to control code generation and subsequent control code deployment. 
This form of control code generation and deployment occurs 
in a setting where non-conscious events can trigger
control code activation and subsequent effectuation while real time control code modification and maintenance is nearly
as difficult as it has become in modern computing practice.
\subsection{Internalized promise issuing (IPI)}
One may contemplate an internalized version of promise issuing where the promise outcome takes the form of
a code inside the promising agent.

Now a promise outcome following single agent internal promise preparation and single agent internalized promise issuing
(IPI) constitutes merely a copy of an intention memorized in the way control code for the mind is 
memorized. 

If this account of IPI is satisfactory then it appears that internalized promising 
 is functionally irrelevant in spite of its existence in principle. If a role must be assigned to IPI that may
be its capacity of strengthening the memorization of the intention that was turned into a promise to self.

Of course deciding internally
to issue a promise externally makes very much sense and must be distinguished from IOODT.

\subsection{Conjectural abilities produced by promise theory awareness}
Following \cite{BergstraDV2011b} when authors propose a novel theory related to human activity and performance
an attempt should be made to express the expected advantages that may come 
about for someone in the audience of the authors 
from having that theory at their disposal. According to that guideline in the current context of a theory of 
promise issuing in relation to decision taking a proposal for the bundle of conjectural abilities of that theory must be put forward.

The conceptual framework that \cite{BergstraDV2011b} offers for analyzing the claimed 
usefulness of a theory is to think in terms
of so-called conjectural abilities that an agent is supposed to acquire when becoming aware of the theory under investigation. Here is a survey of these conjectural abilities in the setting of promise theory and its relation with an
OODT approach to decision taking.

\begin{enumerate}
\item Ability to design (during promise preparation) a promise outcome in such a way that an effective mix of the five 
reasoning processes listed in \ref{reasoning5} above is likely to take place upon promise outcome delivery.

\item Ability to assess when a promise is not just a message in that its understanding 
by agents in scope essentially involves
both inspecting and updating an account of that agent's trust in the promiser.

\item Ability to design a thread which combines decision  and promise issuing in a plausible way. 
Ability and preparedness to provide clarification of the
OODT versus IOODT issue when it comes to promises about decisions and decisions about promises.

\item Ability to provide clarification about issues concerning outsourcing, insourcing, backsourcing, follow-up sourcing, 
and outtasking, of decision taking, decision preparation, promise issuing, and promise preparation.
\end{enumerate}

\section{Contrasting promise and decision}
One might think that promise and decision are very close concepts. That seems not to be the case, however. 
Here are some important differences between (OODT, that is EOODT style) decisions and promises.
\begin{description}
\item{\em Tangibility of decision outcomes.}
In comparison with a decision that may show a considerable gap between a decision and the following decision outcome, 
a promise is closer to a promise outcome. Whereas I insist that following 
OODT a decision outcome is tangible, a promise outcome and more specifically  the items in the 
plurality of corresponding promise outcome instances need not be tangible.

\item{\em Role independence of promises.}
Following \cite{Bergstra2011a} the role played by a deciding agent 
constitutes an essential aspect of the decision outcome. Because of that role,
which is another matter than trust altogether, other agents take notice from the 
decision outcome and may be led into action because of its content.

\item{\em Impact bias of decision versus information bias of promise.}
Plausibly a decision impacts the behavior of agents in scope of the decision 
outcome in such a way that compliance with the outcome is sought. 
In contrast  a promise outcome induces agents in scope of other behavior than mentioned in the
promise body. Context dependent reasoning is asked for in order to make use of the fact that the promise has been made.

\item{\em Linear causation for decisions, non-linear causality for promises.} A decision that $\phi$ must hold may
cause some agents to see to it that $\phi$ becomes true. The causality works in a linear fashion, and the (appearance)
 of the decision outcome may rightly be viewed as a cause for its content becoming true. 
 
 In the case of promises
 things work differently. Indeed assuming that issuing a promise gives expression to an 
 intention held by the promiser, that very intention (and not its expression) 
 causes the  promiser's actions (if any) leading to a fulfillment of the intention. 
 The promiser is expected behave in such ways that the promise is kept. A decider, however, is likely not to be 
 expected to make sure by its own actions that the outcome of its decisions are effectuated.
 
 The promise outcome instance as experienced by a 
 promisee (or an agent in scope of the promise) can have causal influence but so to speak its influence is not effective in a linear fashion. The existence of the promise outcome will not cause the promised state of affairs to 
 arise. Instead the promise has its effect  by complementing the expectation that such (the promised state of affairs becomes true) will be the case with the causation  of responses to that state of affairs by agents in scope of the promise.

\item{\em Acting in compliance versus out of pure self-interest.} 
A promisee is not forced into forms of compliance and is entirely free to make use of 
a promise. In contrast a decision outcome may impact agent behavior irrespectively of an agent's perceived self-interest.

\item{\em Jurisdiction of decision outcomes has no match for promise outcomes.} 
The interpretation of a promise outcome by a promisee is comparable with how a decision outcome is viewed by agents who are outside the 
jurisdiction of the decision taker of that decision. 
Such agents expect that the decision will create follow-up actions by other agents  for 
whom the decision is primarily meant, and they may understand 
the decision as if it were replaced by the collective promise of agents within the jurisdiction of the decision to act in 
compliance with the decision outcome.
\end{description}

\subsection{Promissory decisions}
A special case of EOODT arises if the effectuation of the decision outcome calls for subsequent 
action by the decision taker only and exclusively. Such a decision comes close to a promise and I will 
refer to it as a promissory decision.

The distinction between a promissory decision and a promise lies in the ability of promissory 
decisions to be binding for the decision taker, of course depending on the decision taker's role. Thus, while a promise 
cannot create an obligation for a promiser,\footnote{%
Here I assume strong non-obligationism as specified in the appendix. In that view each promise brings with it a promissory
obligation, but a promissory obligation is only a marginal collateral of a promise and is not (or need not be)
characteristic of the promise. In particular the role of a promiser can only contribute to the significance of resulting
obligations though decision taking, in particular the taking of promissory decisions, where the corresponding promise may
be implied and silent.} a decision outcome can embody an obligation for agents to act and in particular
a promissory decision can create an obligation for the decision taker to act.

\subsubsection{Internalized promissory decisions}
A special case of IOODT arises if effectuation of the decision outcome calls for subsequent 
action by the decision taker only and exclusively. Such a decision comes close to a promise and I will 
refer to it as an internalized promissory decision. An internalized promissory decision can create an 
obligation for the decision taker unknown to any other agent.

\subsection{Combining promise issuing and promissory decision taking}
If a promise is issued and at the same time a corresponding promissory decision is taken, that combination may
introduce an obligation for the promiser to act in compliance with the promise which is not caused by promising alone. 

If a promiser takes an internalized promissory decision concurrently with a corresponding promise, then the promiser may
be subject to an obligation from inside to act in correspondence with the promise.

\section{Decision and promise as manifestations of will}
Questions about free will are complicated by unclarities about whether or not one may or even must assume that the 
world in general, and the working of human brains in particular, is indeterminate. I will assume that (i) the question of determinacy
cannot be answered with certainty and that options for both answers should best be left open, 
(ii) that explanations involving indeterminacy can be viewed as efficient
and abstract stories about a deterministic world (for a deterministically biased observer), and that 
(iii) deterministic treatments 
can be thought of as thought experiments in an indeterminate world (for an observer  rejecting determinism). 

The question of (in)determinacy being unresolvable, or at least assumed to be unresolvable,  one is led to the 
conclusion that the conceptualization of will and free will must be such  that both will and free will 
make sense in a deterministic world
as much as in an indeterminate world.\footnote{%
This is a compatibilist position in that free will and determinism are assumed to be compatible.}
I will assume that free will is manifest if a will is manifest and if that will is in addition free. 
In this paper I will focus on the possible relations between willing, deciding, and promising while
leaving the matter of freeness aside following 
the well-known account of H.G. Frankfurt (\cite{Frankfurt1971}) for that matter.

\subsection{Willing in general and willing to drink in particular}
Willing is a conscious process or perhaps more precisely the conscious layer of a process that involves 
both conscious and unconscious aspects. For an animal the awareness that it 
wants to drink  is a sign that a plan must be made.
Planning is conscious because plans are not mere repetitions of previous action.

In as far as thirst is the expression of a will to eat, that will and the consciousness of it are unrelated to concepts of freedom.
The notion of free thirst seems to be both speculative and nearly irrelevant to the underlying notion of hunger. A
comparison with willing is attractive: willing is a consciousness about a state of mind for which freeness is both speculative and 
nearly irrelevant. The will to drink is a dedicated will, just  the will to take a rest or to search for food.\footnote{%
The will to breath is unconscious. That is plausible because a land based mammal in need of breathing is not in the 
position to follow complex plans to achieve that goal. This is a different matter for a whale who may become conscious of the
need to plan the intake of fresh oxygen in a different and more structured manner than say a human being is likely to do.}

A speculative evolutionary perspective on hunger may be that an organism by having hunger cast as a will to eat
under its conscious processes is more competitive than a organism would be that follows an unconscious reaction pattern.

\subsection{Functional relevance  of willing}
Suppose that a living agent A becomes conscious of its will to achieve objective X. This consciousness may indicate to A
that complex new plans may be needed and that, most important for willing, dedicated forms of 
willing such as the will to have food, may need to be put aside to give room for the realization of a plan 
P that is likely to help A in achieving X.

\section{Promises and decisions in specific contexts}
In this Section I will provide a catalogue of examples where promises and decisions can be applied alternatively or in combination.
The variety of somewhat different promises related to the same issue is rich. One may wonder if this complexity is merely an artifact
of unnecessarily speaking of decisions and promises in the same context or if the 
various distinctions may practical sense, thereby establishing
beyond doubt that promise and decisions are unavoidable concepts. I favor the latter view and I consider the many nuances
that are enabled when speaking of promises and decisions at the same time facilitate essential clarification.

\subsection{Promises and decisions on instruction sequence effectuation}
Th use of an inseq (instruction sequence) X by an agent A consists of the effectuation of 
X (see \cite{Bergstra2012g0}) with an objective in mind 
for which said effectuation is deemed helpful. The objective must be such that if Y is semantically equivalent with X, and 
if moreover Y produces its meaning with comparable computational cost, 
then effectuation of Y may replace effectuation of X in the context of an attempt to the objective.

\subsubsection{Promises about inseq construction and validation}
\begin{enumerate}
\item A promises B that A will produce a satisfactory (for A) set of requirements R for an inseq that A expects to need in the future.
\item A promises B that R constitutes a valid set of requirements for an inseq that it needs to avail of
 in the future 
 \item A promises B that it contemplates offering B the assignment to construct (as a service for A) an inseq X compliant with R,
 under the condition that B accepts validation of the inseq product by a third party that has been selected by A.
 \item B promises A that B will accept the offer made by A, when B is offered (by A) 
 the task to produce an inseq X compliant with R  under conditions C$_P$.
 \item E promises A that (i) it considers R to be a requirement specification of kind K$_R$,  
 (ii) that E has ample experience with validating inseqs that have been produced by other agents against 
 requirements of kind K$_R$, and 
 (iii) that agents A$_1$, A$_2$ and A$_3$ are able and willing to make testimony to that fact.
 \item A promises E that after completion of the production of X by B (for A), the task to validate compliance of X with 
 R will be offered to E against conditions C$_V$.
 \item E promises A that it will accept the offer to construct X against conditions R.
 \item A offers B the task to produce inseq X that meets requirements R under conditions C$_P$.
\end{enumerate}

\subsubsection{Promises about inseq usage and validation}
Here is a sample of promises (promise types) related to the validation and usage of inseqs.
\begin{enumerate}
\item A promises B to use (that is: that A will use, or that A will apply), from now on until recall, inseq X in context C for purpose P.
\item A promises B to deploy (install) inseq X in context C for purpose P.
\item B promises A that it will make use of A's promise to apply P in context C. 
\item A promises B that X complies with requirements R.
\item B promises A that it will make use of A's promise to B about the compliance of X with requirements R.
\item A promises B that X constitutes a fault-free implementation of requirements R.
\item A promises B that X constitutes a defect-free implementation of requirements R.
\item A promises B that $d_e$ constitutes a description of an effectuation $e$ of X in context C
which ends in a failure with respect to requirements R.\footnote{%
In this case X shows a defect against requirements R. Following \cite{Bergstra2012g,Bergstra2012h} the 
existence of a defect of X w.r.t. R 
does not imply the existence of a fault in X.}
\item A promises B that it will test in conditions C$^t$ for the compliance of X with requirements R.
\item A promises B that $d^t_e$ constitutes a description of an effectuation $e$ of X in context C$^t$ 
which ends in a failure with respect to requirements R.  (That is: $d^t_e$ is a log of a failed test for X).
\item A promises B that it has successfully tested for the compliance of X with requirements R.
\item A promises B that it has proven that X complies with requirements R.
\item A promises B that $p$ is a proof that X complies with requirements R.
\item A promises B that  (i) X has been designed so as to comply requirements R by a production team PT,
(ii) with team  experience TE, (iii) working 
with engineering method EM, involving product validation method PVM and and 
(iv) (production) process quality control method PQCM.
\item A promises B to design and develop X in satisfaction of requirements R  within budget $b$ and with help of H.
\end{enumerate}

\subsubsection{Decisions about inseq usage and validation}
\begin{enumerate}
\item A decides that A will use X in context C$_P$ for purpose P.
\item A decides that B is permitted to use X in context C$_P$ for purpose P.
\item A decides that B will use  X and only X in context C$_P$ for purpose P.
\item A decides that it has been given sufficient evidence that ``X complies with requirements R''.
\item A decides that X complies with R.
\item A decides that it has received sufficient evidence from B  that ``X complies with requirements R''.
\item A decides that B will asked to make an offer\footnote{%
An offer is a conditional promise, where the condition states that the offer is accepted. Accepting an offer
consists of promising to make use of the promise contained in the offer.} 
to produce an X in compliance with requirements R.
\item B decides that it will outsource part of the production process for inseqs compliant with requirements of kind R  
to external party  U.
\end{enumerate}
In each of these cases two interpretations of the situation compete for the disambiguation of the scene at hand, 
the EOODT interpretation an the IOODT interpretation.\footnote{%
These extremely simple descriptions are best understood as prose that explains a story about a potentially
 fictional course of events.
If the story is ``real'' that very understanding transpires to the participants of the discussion by ways not formalized in
these descriptions. Here fiction science as meant in \cite{Bergstra2012f} enters the picture.}
 In the EOODT interpretation of the situation requires that A (which may be a person or an informal body,
or some formal committee) produces a decision outcome stating that A will use X in context C$_P$ for purpose P.

The role of A gives authority to the record of this fact and, form there a variety of agents involved in 
the matter may proceed with further action depending on the decision that has ben taken concerning X and its use.

The IOODT interpretation of this decision requires that A is a person (or perhaps an artificial agent) who by deciding
installs in his/her/its mind control a code CC$_{u,X}$ to the extent that A's subsequent actions 
concerning the use of A are consistent with (and likely to be predicted by) the the behavior that would be 
expected from A had A made a commitment to the decision by way of producing a recorded outcome of it.
\subsubsection{Decisions about promises on usage and validation}
\begin{enumerate}
\item A decides that B will be promised (on behalf of A) that use by B of inseq X under conditions C$_P$ is permitted.
\item A decides that A will promise B that use by B of inseq X under conditions C$_P$ is permitted.
\item A decides that B will be promised (on behalf of A) that X constitutes a fault-free implementation of requirements R.
\item A decides that A will promise B that X constitutes a fault-free implementation of requirements R.
\end{enumerate}
Concerning the above four decision descriptions the following can be said. Rather than distinguishing an 
EOODT reading and an IOODT reading of these decision descriptions which is definitely possible in 
each of these four cases, I suggest that the first and third case suggest an EOODT reading where some other agent (say D)
who has not been explicitly mentioned in the text fragment, is supposed to issue the promise to B after  
the decision has been taken and that D's promise issuing is caused (triggered) by D's inspection of the decision outcome
that was produced by A's decision taking.
\begin{enumerate}
\item A decides that A will promise B that $d_e$ constitutes a description of an effectuation $e$ of 
X in context C$_P$ which terminates at a failure with respect to requirements R. 

(Both an EOODT reading and an IOODT reading are plausible in this case. 
If IOODT must be excluded explicit mention of a decision outcome must be made,
if EOODT is to be excluded explicit mention of the absence of an external (to A) decision outcome must be made).

\item A decides that it will promise B that X constitutes a fault-free implementation of 
requirements R in spite of A's knowledge to the contrary. 

(Here ``it'' is read as A and an IOODT reading is plausible).
\item A promises to B that an effort is made by A to decide that B must develop inseq X in compliance with requirements R.

(Promises about decisions are likely to be about decisions in an EOODT sense).
\end{enumerate}

\subsection{EOODT reading versus IOODT reading}
In subsequent examples about scenes with promises and decisions, an EOODT reading 
of the scene will be taken to be the preferred reading giving way to an IOODT 
reading of ``decision taking'' only if the latter reading is clearly more plausible. The
justification of this preference lies in the idea that IOODT readings of decision taking descriptions 
are hardly capable of being understood as recordings of observable facts, whereas EOODT readings can be understood
in observational terms.

If A (EOODT style) decides that A will promise B that P (holds, or will come about, or will be performed), 
then it is plausible that
the promise (that must be issued acceding to A's decision outcome) is issued by an agent D different from both 
A and B where D's action is triggered by becoming aware of the decision outcome reduced by A.

The consequence of taking the above OODT reading of A's decision seriously is that it must be assumed that issuing 
a promise by A can be outtasked (in this case by A to another agent D who is supposed to be acting on behalf of A).\footnote{%
In \cite{B?} I have investigated to what extent decision taking can be outsourced or outtasked, 
leading to the conclusion that outtalking decision taking is plausible while outsourcing decision taking is not. For
terminology on outsourcing and outtasking I refer to \cite{BergstraDV2011c}.}

\subsection{Promises and decisions about informational money transfer}
I refer to \cite{BergstraL2013a,BergstraL2013b} for information about P2P informational monies. 
Using the terminology of those papers the following promises and decisions make sense concerning a P2P informational money IM$_L$.
\begin{enumerate}
\item A decides that  account $a$ is a valid account for IM$_L$.
\item A decides that A has exclusive control of account $a$.
\item A promises B that A has exclusive control of account $a$.
\item A promises B that  A considers account $b$ as being a valid account for IM$_L$ which is under exclusive control of B.
\item B promises A that B has exclusive control of account $b$ and that B is willing to accept transfers to $b$ as payments,
in compensation for transactions of kind K$_t$.
\item A promises B that A will make use of B's promise that ``it has exclusive control of account $b$ 
and that it is willing to accept transfers to $b$ as payments''.
\item A decides that A will transfer amount $q$  from account $a$ to account $b$.
\item A decides that A will promise B that A will transfer amount $q$  from account $a$ to account $b$.
\item A promises B that A will transfer amount $q$  from account $a$ to account $b$.
\item A promises B that A has posted a transfer of an amount $q$  from account $a$ to account $b$ on the P2P network for IM$_L$.
\item A promises B that A is not attempting to launch a double-spending attack involving transfers from $a$ to $b$.
\item B promises A that B will wait for the transfer amount $q$  from account $a$ to account $b$.
\item B promises A that B will acknowledge to A  the successful transfer of an amount $q$  from account $a$ to account $b$.
\item B decides that a transfer of an amount $q$  from account $a$ to account $b$ has been successfully performed.
\item B decides that it will promise A to consider the  transfer of an amount $q$  from account $a$ to account $b$ 
to have  been successfully performed.
\item B promises A that it  considers the  transfer of an amount $q$  from account $a$ to account $b$ 
to have  been successfully performed.\footnote{%
In a Bitcoin like system (see \cite{Nakamoto2008}) B cannot be 100\% sure of this judgement and B 
must issue the promise on the basis of satisfactory but potentially non-conclusive evidence.}
\item A decides that the  transfer of an amount $q$  from account $a$ to account $b$ 
has  been successfully performed.
\end{enumerate}

\subsection{Budget proposals and budget announcements}
As mathematical objects representing a budget for an organization or an activity, say Q, pertaining to a certain timeframe 
I will use tuplices from tuplix calculus (TC) as defined in \cite{BergstraPZ2008}. 

In its simplest form a tuplix specifies an interface and for each element (label) of the
interface a positive amount if money is flowing in (income from the perspective of Q) and a negative amount if money is flowing
out (expenses). The elements of the interface represent cost categories (cost labels with negative entries expected) 
and result categories (result labels with
positive entries expected). TC allows for free meadow variables typically working in the ordered meadow of 
rational numbers (see \cite{BergstraBP2013}). A tuplix with free variables can represent an architecture for a budget which admits various instantiations.

Further TC allows hiding or binding of free variables thus allowing for more complex relations between the various label entries. 

Let t$_Q$ be a tuplix that specifies a generic budget for Q, written by agent A who is responsible for managing activity Q
inside an organization managed by agent B.  The tuplix t$_Q$ specifies advance of the timeframe a family of 
instantiations that each represent possible budgets for the activity Q when actually realized. Parameters of t$_Q$ 
may represent say numbers of participants, costs of certain services provided to participants, 
volumes of resources claimed for duration of Q and so on.

\subsubsection{Promises and decisions concerning t$_{\bf Q}$}
\begin{enumerate}
\item A decides that Q can be carried out with a yet unspecified instantiation (refinement)
of t$_Q$ as its operational budget.\footnote{%
That conclusion provides (at least in the eyes of A) a lower bound
for the result of the activity (or for its loss if the result is negative). It may also provide an upper 
bound for the profit that may be realized through Q.}

\item A promises B that Q can be carried out with an instantiation of t$_Q$ as its operational budget.
\item B decides that B will only accept (as a budget for Q) instances of t$_Q$ that are instances of a specific
 instance (substitution result) $\sigma_B$(t$_Q$).
\item B promises A that B will accept  $\sigma_B$(t$_Q$) as a generic budget for Q.
\item A decides that Q can be carried out within an instance of budget $\sigma_B$(t$_Q$).
\item A promises Q that A can carry out Q within budget $\sigma_B$(t$_Q$).
\item B promises A that B accepts A's promise to organize Q within the constraints of budget  $\sigma_B$(t$_Q$).
\item A decides to organize Q in conformance with the requirements formulated by B.
\item Some time later, say at time $r$ before the time frame for Q will begin, A decides  that a further refinement of the budget,
say $\sigma_r$($\sigma_B$(t$_Q$)) can be made on the basis of incrementally available figurers about how Q will unfold.
\item A decides to issue the promise to B that Q will be effected within the constrains of budget $\sigma_r$($\sigma_B$(t$_Q$)).
\item A promises B that Q will be carried out within budget $\sigma_r$($\sigma_B$(t$_Q$)).
\item B promises A that when Q is carried out the final account may show a result that is $S_r$ (additional guarantee promised
by B at time $r$) short of the result predicted  by budget $\sigma_r$($\sigma_B$(t$_Q$)).
\end{enumerate}

\subsection{Promises and decisions for avoiding multi-valued logic creep}
Consider the following promises about a rational value $X$ that A and B both know to exist.\footnote{%
I acknowledge 
the help of Alban Ponse with developing this paragraph on division by zero.} 
The promised state of affairs is supposed to hold just after the promise has beent issued and is 
supposed to persist for enough time to enable B several consecutive inspections of the value of $X$: 
\begin{enumerate}
\item A promises B that $0 \leq X \leq 2$ and $0 \leq \frac{X }{ X - 1} \leq 2$.
\item A promises B that $0 \leq X \leq 2$ and  $0 < \frac{X }{ X - 1} < 2$.
\item A promises B that $\frac{X}{X} = 1$.
\item A promises B that $\frac{X}{X} \neq 1$.
\end{enumerate}

\subsubsection{What this means in the meadow of rational numbers}
Working in an ordered meadow~\cite{BergstraBP2013}, in particular the meadow of rational numbers (as specified in
\cite{BergstraT2007}, and more succinctly in \cite{BergstraM2011a}), 
these promises can be simplified to the following forms:
\begin{enumerate}
\item A promises B that $X \in \{0,1,2\}$.
\item A promises B that $X = 1$.
\item A promises B that $X \neq 0$.
\item A promises B that $X = 0$.
\end{enumerate}
An agent in scope of these promises may be unaware of the virtues of meadows, 
or may be unimpressed by the mathematics  of meadows, and may prefer doing 
elementary mathematics in a more conventional manner. That agent
must take worries about division by zero into account while not taking such worries too seriously at the same time.

\subsubsection{A conventional approach}
To an agent who is mildly worried about division by zero related issues it may be suggested that 
the initial five promise descriptions require some clarification and that 
after applying due care rewriting the respective promise bodies is possible and the following 
promises are meant instead. In the following  rewritten promises the body is augmented 
with a first component that expresses a condition which guarantees that the second 
component avoids division by zero.
\begin{enumerate}
\item A promises B that $0 \leq X \leq 2$ and $X \neq  1$ and $0 \leq \frac{X }{ X - 1} \leq 2$.
\item A promises B that $0 \leq X \leq 2$ and $X \neq  1$ and $0 < \frac{X }{ X - 1} < 2$.
\item A promises B that $X \neq 0$ and $\frac{X}{X} = 1$.
\item A promises B that $X \neq 0$ and $\frac{X}{X} \neq 1$.
\end{enumerate}
According to conventional wisdom, somehow based on the intuition behind short-circuit logic,  
one makes the tacit assumption that
in ``$\Phi$ and  $\Psi$'', once $\Phi$ is false (or rather turns out to be false) $\Psi$ need not be evaluated, 
and that in that case and for that reason even the form of $\Psi$ can  be ignored. With this reading instruction in mind
 the original five promises can be simplified to as follows.
\begin{enumerate}
\item A promises B that $0 \leq X \leq 2$ and $X \neq  1$ and $X \in \{0,2\}$.
\item A promises B that $0 \leq X \leq 2$ and $X \neq  1$ and $\bot$.
\item A promises B that $X \neq 0$ and $\top$.
\item A promises B that $X \neq 0$ and $\bot$.
\end{enumerate}
Justifying these simplifications by means of calculational rules or formal argument is another matter altogether. Why
would 
``$0 \leq X \leq 2$ and $X \neq  1$ and $0 \leq \frac{X }{ X - 1} \leq 2$''
 be equivalent to 
``$0 \leq X \leq 2$ and $X \neq  1$ and $X \in \{0,2\}$''? 
Formalizing this equivalence as
``$(0 \leq X \leq 2 \wedge X \neq  1 \wedge 0 \leq \frac{X }{ X - 1} \leq 2) \leftrightarrow (0 \leq X \leq 2 \wedge X \neq  1 \wedge X \in \{0,2\})$'' requires a three-valued logic or another non-classical logic
in case one insists not to read division as a total function (the approach taken in meadows).

Without taking this formalization on board it cannot be denied that some difference between ``$\Phi$ and $\Psi$'' 
and ``$\Phi \wedge \Psi$'' must be taken into consideration.

Introducing that sort of distinction requires, by way of a logical point of departure, either the use of 
short-circuit logic, which implicitly involves the use of an order of evaluation as an abstraction of the progression  of time,
or the use of some multi-valued logic. Multi-valued logic entering the picture without 
deliberately having been taken on board is a phenomenon that I will speak of as multi-valued logic creep. The relative 
 ``difficulty'' of multi-valued logic in comparison with the remarkable simplicity of two-valued logic for the purpose of 
 describing number systems and reasoning about those has been highlighted in \cite{BergstraM2011a}.

\subsubsection{Multi-valued logic creep (MVL creep)}
The difficulties caused by MVL creep are twofold, firstly one must be very precise about which multi-valued logic 
one will be using,  and secondly the notational details, semantics, and and proof systems, for multi-valued logic 
approaches to elementary mathematics are both uncommon and somehow non-trivial, if not counterintuitive. 
The presence of MVL creep implies no less than a spark of puzzling informality in the midst of a 
collection of seemingly precise assertions. MVL creep may render a topic unnecessarily obscure.\footnote{%
And the misconception may arise that MVL creep is caused by the interaction between formalized mathematics 
and the natural language in which it is embedded. To the contrary, MVL creep arises exclusively because of particularities of
the design of the mathematical structures at hand.}

\subsubsection{Promising to avoid MVL creep}
Meadows carry an implicit promise:  the implicit promise (see~\cite{BergstraB2008}) of 
meadows is that the use of meadows facilitates avoiding MVL creep in elementary mathematics. 

Relying on this implicit promise an agent A may produce the explicit promise that
``in the coming time I will interpret rational expressions with rational number variables as being meant over the ordered 
meadow of rational numbers''. Issuing the latter promise may be based on a decision taken to that extent: A may decide
(either EOODT style or EOODT style) that
``in the coming period I will work with rationals within the meadow of rational numbers only''. Assuming that A has the authority
to take this decision and assuming that A interacts with agents B who are initially unaware of A's dislike of MVL creep and of
A's preference for meadows as a method for avoiding MVL creep, it is plausible that
whenever a new interaction with an agent B starts, A promises that ``meadows will be used'' 
and that MVL creep will be avoided.\footnote{%
That is more easily said than done, however, and here lies the advantage of thinking in terms of promises in such cases.
If MVL creep occurs nonetheless, the promises downgrades his or her trust in the promiser's ability to keep that
particular promise.}

In this case the gap between decision making and decision taking is apparent. In the decision making process preceding
the decision taken by A ``to use meadows'' an extensive decision preparation phase  may occur during which
 meadows are studied and the various pro's and con's of meadows are evaluated.

\subsubsection{Silent decisions that come with promising the use of meadows}
If A uses arithmetic in terms of meadows the risk arises that other agents processing  A's calculations fail to notice
that some calculations lead to results on the basis of evaluation of division by zero to zero while the result of that calculation
admits no meaningful application. A is assumed to silently promise (see \ref{SilentPr}) that it will warn agents in A's audience explicitly when
such risks arise. Such warnings may be omitted if A promises to comply with an appropriate relevant division 
convention (see \cite{BergstraM2011a}).\footnote{The relevant division convention is a special case of the relevant use convention as proposed in \cite{BergstraM2012}. The relevant division convention and its generalized form of the relevant use convention have been introduced with
the expectation to constitute a preferred way of looking at  partial functions.}
\section{Concluding remarks}
A satisfactory match between outcome oriented decision taking and autonomous agent based promising has been developed. 
When used in combination decision taking and promise issuing allow a 
flexible language for describing aspects of elementary technology. 
The later has been exemplified in the contexts of informational money transfer, 
instruction sequence effectuation, budgeting design and announcement, and division by zero.

\appendix
\section{Aligning SToP terminology  to OODT terminology}
I will provide a summary of the account of promises given in SToP (\cite{BergstraB2008}).
In order to achieve an alignment between the OODT terminology and SToP terminology
the latter is adapted to the former in ways consistent with the main text of the current paper.

SToP approaches the
subject of decision taking at a particular level of abstraction. In SToP aspects of timing and causality play almost no role. 
As a consequence, the SToP definition of a promise (and related
concepts) must be understood as being meant to pertain only, or at least primarily, 
to that specific level of abstraction.\footnote{%
This dependency of concept definitions from the specific abstraction level at hand is very visible in the theory
of information systems where often one will speak of an employee if no more than a reference to a 
bundle of (potentially) personal data is meant. Of course it is not meant that the concept of an employee 
is exhaustively defined, and for that reason equally useful for other applications, by that kind of specification for it.}

In SToP a promise combines the action of promise issuing an the promise outcome together
with promiser, promisee, and so-called scope in a single entity called promise. Aligning the terminology
of SToP to that of OODT does not affect the differences in abstraction level, 
these persist in a modified terminology just as well.

By adapting SToP terminology to
an alignment with OODT one obtains what might be termed OOToP (outcome oriented theory of promises) in which
every occurrence of promise must be disambiguated by either speaking of promise issuing, of promise outcome,
or of promise preparation. When referring to promise as meant in SToP (that is the abstraction as aimed for in SToP
but otherwise writing in adapted terminology nevertheless) I will write ``promise (in SToP)''.

\subsection{Implicit promises versus explicit promises}
A very frequent usage of the term promise is in phrases
like ``the weather is promising'' or ``the promise of quantum computing''.
These are  promises are implicit because  the the existence of a promiser 
is not assumed and no information transfer is implied.

An explicit promise comes about from an action performed
by a promiser in an appropriate context. OOToP exclusively focuses on
explicit promises, for that reason it is assumed by default that promises are explicit.

\subsection{Silent promises versus explicitcit promises}
\label{SilentPr}
Not mentioned in \cite{BergstraB2008} is the notion of a silent promise, that is a promise which can be
assumed to have been implicitly issued by a promiser given a context. For example by allowing clients into 
a restaurant a silent promise is issued by its staff that the guests are welcome. The main difference between a silent 
promise and an implicit promise is that a silent promise is issued by an agent (the promiser) and the event
of issuing takes place in space and time. For implicit promises not even a hypothetical promiser comes ito play.

\subsection{Constituents of a promise in SToP}
A promise (in SToP) consists of the following components.
\begin{itemize}
\item There is a promiser (or source agent).
\item There is a promisee (or recipient agent) which might be the same as the source.
\item There is a body which describes the nature or content of the promise.
\item There is a scope including the promisee of agents who will be made aware of the promise.
\item The body specifies a quality (what kind of action or state of affairs is promised) and a quantity (how late, how 
long, how many, how often, how expensive etc.)
\item  Some form of documentation\footnote{%
Documentation is meant in a flexible manner. It may range from formal signed and carefully 
distributed physical documents to the impression left in the minds of a group of agents after
having seen or heard the promiser  issuing the promise.}
serving as a carrier for information transport between promiser, promisee, and other agents in scope of the promise.
\end{itemize}

A promise comes about from a {\em possible intention}.\footnote{%
The term intention should be read quite liberally. It also refers to a state of affairs that is intended to hold,
or to a strong expectation that one intends to have been generated with sound justification.}  Possible intentions are both
temporally and logically prior to promises.  By being made public (even unintentionally) an intention becomes a promise. 

\subsection{Promise outcomes are documented apparent intentions}

In the realm of all possible formulations about agent behaviour 
the concept of an intention stands out as an important foundation.

\begin{definition}[Current intention of an agent $A$]
A current intention of an agent $A$ is description of a possible behaviour, 
or goal, or objective, or state of affairs,
that is contemplated by $A$ with the understanding that it can be and 
preferably (for $A$) will  be brought to realization.
\end{definition}

\begin{definition}[Possible intention for an agent $A$]
A possible intention for an agent $A$ is a description of a possible behaviour, 
or goal, or objective, or state of affairs,
that may but need not currently (at the time of qualifying the description) be 
contemplated or preferably brought to realization
by $A$, and which might be in some (possibly different) 
circumstances a current intention of $A$.
\end{definition}

Obviously a current intention is also a possible intention. 
But if an utterance of $A$ announces a possible intention that
need not be a current intention, it may only appear to be a current intention. 

The components of an intention are as follows: a source agent who formulates the
intention, a target agent if the intention is directed at a potential
subject, and a body which explains the quality and quantity of the
intention. Only the source of an intention
knows about the intention, i.e. the scope of an intention is the source only.
There are no witnesses.

The set of all possible intentions should be distinguished from actual
instances of intentions selected by an particular agent. We shall
sometimes use the phrase ``possible intentions'' to mean this full set
of abstract entities to emphasize when we wish to signify a general
description of behaviour rather than an individual
agent's choice.

\begin{definition}[Commitment]
Commitments are current intentions that we are committed to. We may call them
intended intentions, or equivalently real intentions, intentions that
we hold, or committed intentions. The commitment of an intention exceeds its 
merely being current
in that it is stable and persists in time until some achievement of the intention
will take place or until some overruling considerations invalidate the commitment.
\end{definition}

When passing by a shop one may feel a current intention to buy a nice gadget, only
to be relieved of that intention (or rather its currency) after noticing its price. If however,
the price is quite good, but the shop is closed at the time of passing along, then a current intention
to buy the same item can become activated with the status of a commitment, only to be 
terminated when the item has been acquired or when unexpected problems turn out to stand in the way.

\begin{definition}[Intention utterance]
An agent $a$ produces an intention utterance if $A$ produces an expression of a description of a possible intention.
\end{definition}

What matters for our discussion on promises is intention utterances that seem to be real.
That leads to the idea of an apparent intention utterance.

\begin{definition}[Apparent Intention utterance]
An utterance expressing a possible intention (of a principal agent) with the 
contextual appearance of an intention. Apparent intentions, 
may be drawn from the following range:
\begin{description}
\item [\em Real intention:] (alternatively: commitment, true intention, or intended intention) what is announced 
corresponds to wha the agent expects that will happen, or that (s)he will do,
or what holds or what will hold. 

In other words the apparent intention is real if it is a commitment (and therefore current).

\item [\em Incidental intention:] (alternatively: non-committing current intention) what is announced 
corresponds to why the agent expects that will happen, or that (s)he will do,
or what holds or what will hold, but only a the time of expression. 

\item [\em Indifferent intention:] (alternatively: quasi-intention) the issuer has no current intention corresponding to the
utterance, and no current conflicting intention either. 

An indifferent intention
is currently contemplated as a possible behaviour, goal, objective, or state of affairs, but its bringing about
is not preferred, and thus an indifferent intention is not a current intention.

\item [\em Deceptive intention:] (also: misleading intention) the announcement might seem to be real for an audience in 
scope but it is a lie from the perspective of the promiser.

A deceptive intention is incompatible regarding realization with a current intention, though this may be only
known to the principal agent.

\item [\em Invalid intention:] (alternatively: manifest lie) all observers may notice a discrepancy 
between what is announced and the facts.

The invalidity of an invalid intention will become clear to agents in scope of that utterance.
\end{description}
\end{definition} 

The idea of an apparent intention is that at face value it is like an intention from the perspective of an
observer but there is a considerable degree of freedom in connection with a so-called underlying intention, 
the existence of which we postulate in the following definition.

\begin{definition}[Underlying intention (of an apparent intention utterance)]
Given an apparent intention utterance of an agent, there is an underlying 
intention (which need not be comprised in the same utterance) as well. We will distinguish five cases, 
corresponding to the case distinction
of intention utterances:
\begin{description}
\item [Real intention:]  The underlying intention of a real intention is that same intention.
\item [Incidental intention:]  The underlying intention of an incidental intention is that same intention which
is known to be coronet but non-committing.
\item [Indifferent intention:] The underlying intention of an indifferent intention is empty.
\item [Deceptive intention:] The underlying intention of a deceptive 
intention differs significantly from the (deceptive) intention.
\item [Invalid intention:] The underlying intention of an invalid 
intention differs noticeably (for observing agents) from the (invalid) intention.
\end{description}

\end{definition}

We will assume the agents keep underlying intentions private. 
Otherwise new levels of complexity emerge as underlying intentions may 
turn out to split over the same distinction of four cases recursively.

\begin{definition}[Promise (in SToP)]
A promise is an apparent  intention  of an agent, 
(the promiser or promising agent) 
the utterance of which has been documented (as a promise outcome)
within a scope that goes beyond the promiser.

According to the definition of intention utterances, a promise (in SToP)  brings
with it an apparent intention and an underlying intention, and five cases can be 
distinguished for promises (in SToP): real, incidental, indifferent, deceptive, and invalid.
\end{definition}

\begin{definition}[Promissory obligation]
With each promise  of an agent $A$ an obligation is connected, the so-called 
promissory obligation. It is that obligation to which the agent has become
obliged by making the promise.
\end{definition}

Promissory obligations are an important tenet of the philosophy of promises, and
SToP does not deny their existence. However SToP opposes to what it refers to as
obligationism.

\begin{definition}[Obligationism]
Obligationism refers to the viewpoint that (i) promises are characterized by
a unique capacity to (auto)generate an obligation (specifically the promissory obligation) 
for the promising agent, and that (ii) the essence or content of a promise is fully captured by its 
promissory obligation.
\end{definition}

\begin{definition}[Non-obligationism]
Non-obligationism  denotes the belief that obligationism is false.
\end{definition}

\begin{definition}[Strong non-obligationism]
Strong non-obligationism denotes the belief that obligationism is false and that in addition
the concept of promise (in SToP) may be accounted for without making use of the concept of an obligation.
\end{definition}

SToP incorporates a preference for non-obligationism over obligationism, and aims at  a strongly 
non-obligationist account of promises. 

\end{document}